\documentclass[11pt,a4paper]{article}
\usepackage{lineno,hyperref}
\usepackage{amsmath}
\usepackage{amsfonts}
\usepackage{amssymb}
\usepackage{graphicx}
\usepackage{graphics}
\usepackage{multicol}
\usepackage{subfig}
\usepackage{geometry}
\usepackage{xcolor}
\everymath{\displaystyle}
\usepackage{authblk}

\title{Network growth with preferential attachment and without ``rich get richer'' mechanism}

\author[]{A. Lachgar, A. Achahbar \thanks{E-mail address: fattah@uae.ma}}
\affil[]{{Condensed Matter Group, Department of Physics, Faculty of Science, B.P 2121, T{\'e}touan, Morocco}}

\begin{document}
\maketitle
\begin{abstract}

We propose a simple preferential attachment model of growing network using the complementary probability 
of Barab\'asi-Albert (BA) model, i.e., $\Pi(k_i) \propto 1-\frac{k_i}{\sum_j k_j}$. In this network, new nodes are 
preferentially attached  to not well connected nodes. 
Numerical simulations, in perfect agreement with the master equation solution, give an exponential degree distribution. 
This suggests that the power law degree distribution is a consequence of preferential attachment probability together 
with ``rich get richer'' phenomena.\\
We also calculate the average degree of a target node at time t $(<k_s(t)>)$ and its fluctuations, to have a better 
view of the microscopic evolution of the network, and we also compare the results with BA model. 
\end{abstract}

\section{Introduction}

 \hspace{0.5cm}In recent years there is a growing interest to study the evolution of complex networks and to develop models
 that reflect certain properties of the real networks using some 
 statistical mechanics techniques, graph theory and computer simulations \cite{BA,AB,DM,NEW}. One of the most important properties 
 studied in networks is the degree distribution of nodes which is the probability $P(k)$ of a node to have degree 
 $k$. We can distinguish three main laws of degree distribution: Poisson law where
 $P(k)=e^{-<k>}\frac{<k>^k}{k!}$, 
 power law with $P(k)\sim k^{-\gamma}$ and $\gamma$ represents the degree exponent, and exponential law with
 $P(k) \sim e^{-\frac{k}{c}}$ where $c$ is constant.\\
 It appears that in nature most networks follow the last two distribution laws referred to above.
 Barab\'asi-Albert reinvented price's power law degree distribution network by introducing a simplified model based 
on both growth and preferential attachment. The resulting scale-free network is widely observed in variety of systems 
such as publication citation networks, many social networks, protein and gene networks. However, there are other real 
networks that follow an exponential law, for example, Worldwide Marine Transportation Network \cite{DG}, the North American
 Power Grid Network \cite{AA}, neural network of the C.elegans \cite{AY}, and the Email Network at the University of Rovira i 
Virgili (ENURV) in Spain \cite{GD}.\\
The exponential law seems to be the result of growing network by randomly adding new nodes and links. On the other hand
 the power law seems to appear when nodes are added to the network one at time and are linked with nodes already well 
connected.\\ 
Many ideas on the formation of networks have been reviewed in the recent years. For example, 
Barab\'asi, in his earlier work, asserted that the preferential attachment and growth are both required to generate
 a scale-free network \cite{BJ}. Actually, it seems that growth is not necessary for such a purpose \cite{XZ}.
Furthermore, it is intuitive that preferential attachment without rich get richer effect does not generate a 
scale-free network \cite {SA}. Krapivsky et al. \cite {KR} have studied non linear preferential attachement with 
$\Pi(k_i) \propto k_i^{\gamma}$, and they shown that for $\gamma<1$ the mechanism produces a stretched exponential
degree distribution.\\
Despite many efforts, consistent theory of networks in evolution is still lacking and there is not yet a general principle 
predicting the topology of a formed network.\\
Aiming to understand the formation and the evolution of complex networks, many models were introduced to 
investigate the microscopic processes implicated in the resulting network structure.\\
In this context we introduce a simple complex network model growing with linear preferential attachment mechanism and without 
rich get richer effect.The objective is twofold: first to check if the power law degree distribution remain in the absence of 
the rich get richer scenario and, second, to see for eventual microscopic differences between scale-free and homogeneous
 networks. 
\section{Degree distribution} 
\begin{figure}[h]
\centering
\includegraphics{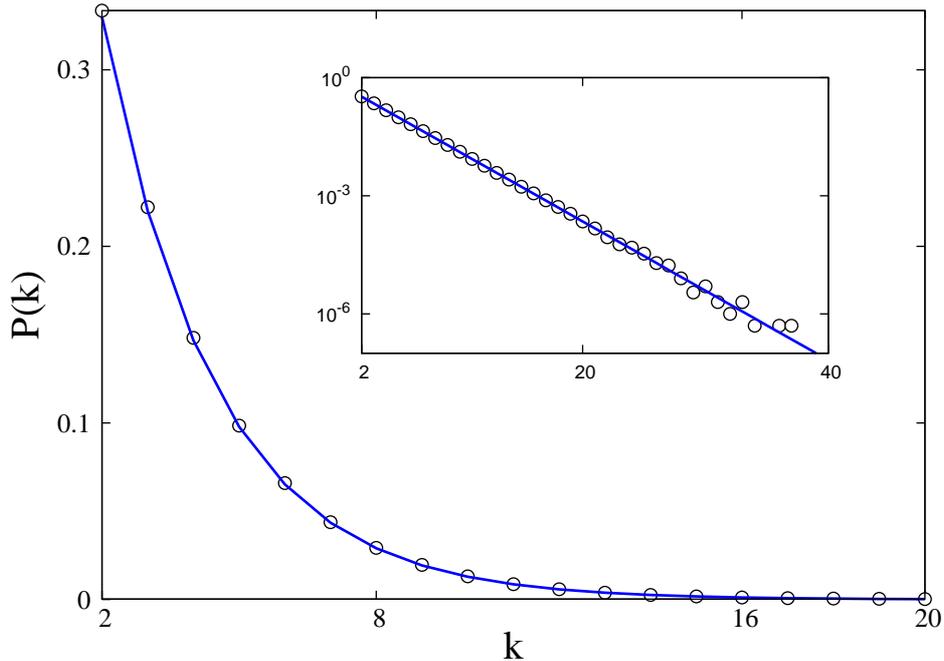}
\caption{Simulation results (circles)  for $n=2.10^6$, $m=2$, $m_0=3$, and numerical solution 
(solid line) of Eq.~\eqref{eq4}. In the inset we plot the same data in the log-linear scale.}
\label{fig1}
\end{figure}

Similarly to the original BA model, our network evolves according to two mechanisms: 
the growth and the preferential attachment.
Nodes entering the network prefer to attach 
to nodes with low degree, then the probability $\Pi(k_i)$ that one of the links of a new node connects to
node i depends on its degree $k_i$ such that $\Pi(k_i)=C\big(1-\frac{k_i}{\sum_j k_j}\big)$, where $C$ is a normalization 
constant.\\
In connection with social networks, if we consider the degree of nodes as describing the wealth of people in a capitalist
society, it is known \cite{PA} that we live in a world where rich get richer, but what kind of society we will have
if there is no favors to rich people, and there is instead a continuous subvention to poor people?.\\ 
To implement our idea, we start with $m_0$ nodes, each one with $m$ links. At every time step we add a new node with $m$ edges 
that link the new node to m different nodes already present in the network.
The probability that the new node is connected to a node  $i$  of degree 
 $k_i$  is $\Pi(k_i)=C\big(1-\frac{k_i}{\sum_j k_j}\big)$.
The normalization constant $C$ is deduced from the condition $\sum_{i=1}^t\Pi(k_i)=1$, which gives  $C=\frac{1}{t+m_0-1}$.
 $t$ is the time when the last node was created and represents also the number of nodes added to the network.\\
For this model, the master equation can be written as: 
\begin{align}
 (t+1)P(k,t+1)=tP(k,t)+m\Pi(k-1,t)tP(k-1,t)-m\Pi(k,t)tP(k,t)+\delta_{k,m},
\label {eq1}
 \end{align}
where $\delta$ is the Kronecker symbol.\\
The corresponding stationary equation takes the form:
\begin{align}
(t+1)P(k)=tP(k)+m\big(1-\dfrac{k-1}{2mt+mm_0}\big)\dfrac{tP(k-1)}{t-1}-m\big(1-\dfrac{k}{2mt+mm_0}\big)\dfrac{tP(k)}{t-1}
+\delta_{k,m},
\label{eq2}
\end{align}
where we used $\sum_j k_j=2mt+mm_0$.
For large time we get
\begin{align}
P(k)&= 
\begin{cases}
\dfrac{2mt - (k-1)}{2t + 2mt - k}P(k-1), \quad \textrm{for }  k>m,\\
\\
\dfrac{2t}{2t + 2mt - m}, \quad\textrm{for }  k=m.
\end{cases}
\end{align}
The above recurrence relation yields the following solution:

\begin{align}
P(k)&= 
\begin{cases}
\dfrac{2t}{2t + 2mt - m}\prod^k_{j=m+1}\left( \dfrac{2mt -j + 1}{2t + 2mt - j}\right), \quad \textrm{for }  k>m,\\
\\
\dfrac{2t}{2t + 2mt - m}, \quad\textrm{for }  k=m.
\end{cases}
\label{eq4}
\end{align}
Although this equation is not in a closed form, numerical estimation of $P(k)$ is straightforward as shown 
in Fig.~\ref{fig1}.\\
We also simulate the network with sizes up to $n=2\times10^6$, initial number of nodes $m_0=3$ and $m=2$. 
The simulation results strongly support the analytical findings (see Fig.~\ref{fig1}).

We observed in simulations that $k$ remains less than $40$ for $t=2.10^6$, we then take $t\gg j$ in Eq.~\eqref{eq4} and
we obtain
\begin{align}
P(k)&\approx 
\begin{cases}
\dfrac{1}{1+m}\Big(\dfrac{m}{1+m}\Big)^{k-m-1}, \quad \textrm{for }  k>m,\\
\\
\dfrac{1}{1+m}, \quad\textrm{for }  k=m.
\end{cases}
\label{eq5}
\end{align}
After normalization we get the exponential degree distribution $P(k)=Ae^{-A(k-m)}$, with $A=\ln(\dfrac{m+1}{m})$.
The inset in Fig.~\ref{fig1} shows the exponential form of $P(k)$ and the excellent agreement between simulations 
and theoretical results.   
This clearly confirms that the preferential attachment alone is not sufficient to produce scale-free networks.

\section{Comparison with BA model}
\begin{figure}[h]
\centering
\includegraphics{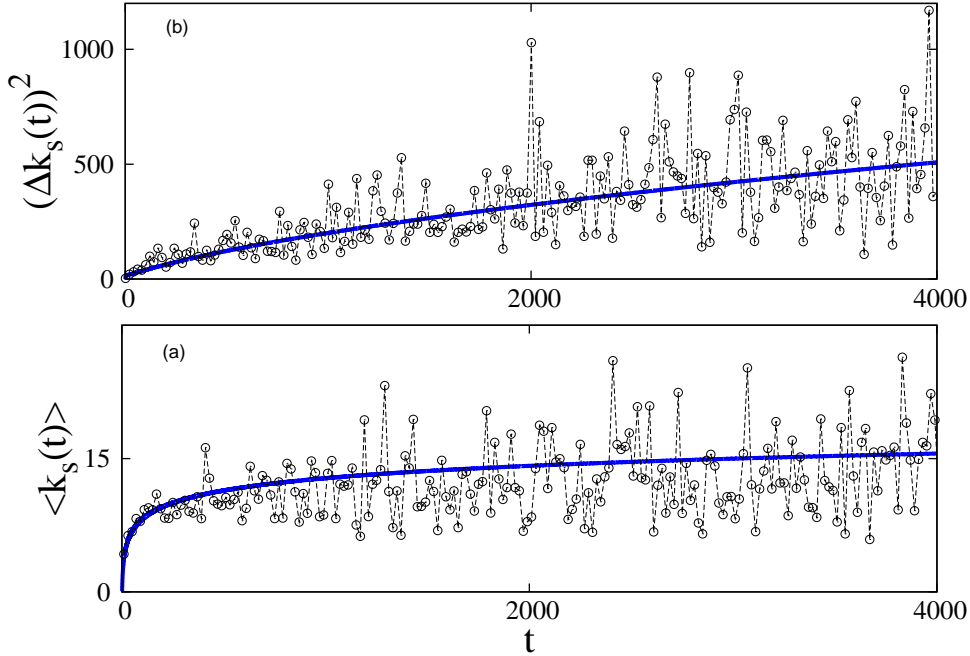}
\caption{(a) Evolution of $<k_s(t)>$ in the BA model, the solid line represents Eq.~\eqref{eq7}. 
(b) Evolution of fluctuations of $<k_s(t)>$, the solid line represents Eq.~\eqref{eq10}. 
Circles joined by dashed lines in both cases are simulations 
data averaged over 20 runs for $m=2$, $m_0=3$.}

\label{fig2}
\end{figure}

We search for differences between heterogeneous and homogeneous networks by comparing our 
model with the BA model.
The degree distribution alone is not enough to characterize networks. Computing others microscopic quantities may
help to have better insight into their evolution and formation. 
It turns out that scale-free network has nodes with important degree (hubs), while random network has no apparent structure. 
Evaluating the instantaneous average degree of target node $<k_s(t)>$ and its fluctuations, can give quantitative
information about hubs in the network. In fact, $<k_s(t)>$ is
somehow related to the instantaneous average degree of hubs, because when choosing randomly nodes, hubs have more chance to
be selected.    \\
Firstly, we analyze $<k_s(t)>$ and $<k^2_s(t)>$ in the BA network
\begin{eqnarray}
<k_s(t)>=\sum_{t_i=1}^t\Pi(k_i)k_i(t)+m_0\Pi(k_0)k_0(t),
\label{eq6}
\end{eqnarray}
where $\Pi(k_i)=\frac{k_i(t)}{2mt+mm_0}$, $t_i$ is the time when the node $i$
was created, and $k_0(t)$ is the degree of initial nodes at time $t$.\\
Solving the mean field equation $\dfrac{\partial k_i(t)}{\partial t}=~ m\Pi(k_i)$, we obtain
$k_i(t)=m\Big(\frac{2t+m_0}{2t_i+m_0}\Big)^{\frac{1}{2}}$.\\
Inserting the last expression in Eq.~\eqref{eq6}, we get  
\begin{eqnarray}
<k_s(t)>&=&m\Big(\sum_{t_i=1}^{t}\dfrac{1}{2t_i+m_0}+1\Big) \\
&=&m \Big(\ln (2t+m_0)+\gamma-a+\frac{1}{2(2t+m_0)}+O(\frac{1}{t^2})\Big),
\label{eq7}
\end{eqnarray}
where $\gamma$ is the Euler constant, and $a=\frac{1}{2}+\frac{1}{3}+\ldots+\frac{1}{1+m_0}$.\\
Good agreement is obtained as shown in Fig.~\ref{fig2}(a) between Eq.~\eqref{eq7} and simulation results even for the
first moments of the evolution. 
$<k_s(t)>$ grows indefinitely with time and diverges for infinite network (or $t \to \infty$) due to the fact that,  
in heterogeneous networks, hubs are more likely to be selected and linked with new nodes.\\ 
On the other side, the average degree of the network remains finite \cite{BAR,CO} since the majority
of nodes have a small degree and the weight of hubs is small.

\begin{figure}[h]
\centering
\includegraphics{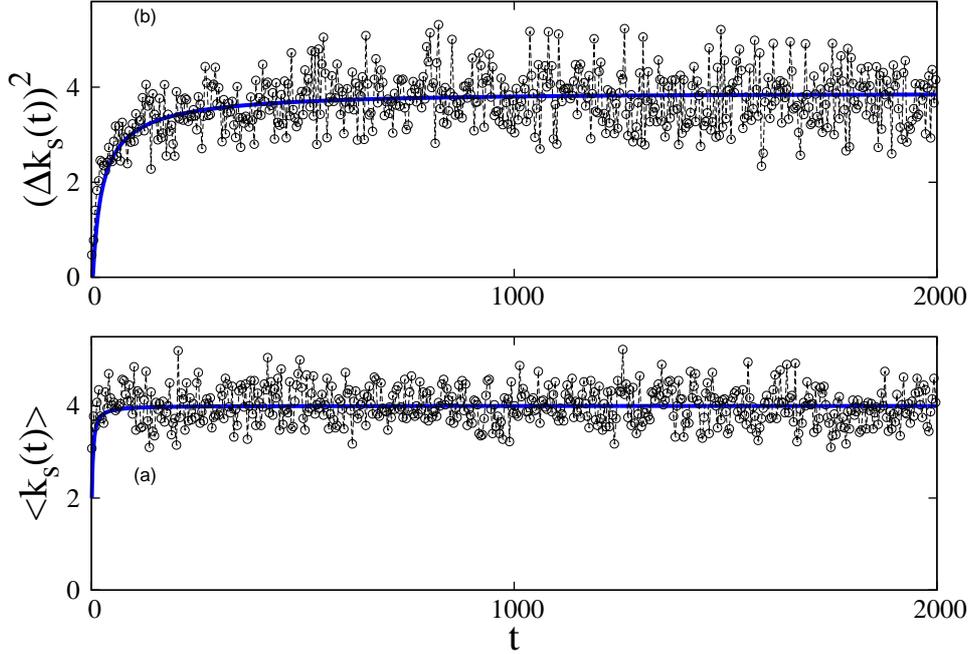}
\caption{(a) Evolution of $<k_s(t)>$ in our model, the solid line represents Eq.~\eqref{kom}. 
(b) Evolution of fluctuations of $<k_s(t)>$, the solid line represents the numerical
solution of Eq.~\eqref{kom} and Eq.~\eqref{k2om}. Circles joined by dashed lines in both cases are simulations 
data averaged over 20 runs for $m=2$, $m_0=3$.}
\label{fig3}
\end{figure}

The second moment $<k^2_s(t)>$ is written as
\begin{align}
<k^2_s(t)>&=\sum_{t_i=1}^t\Pi(k_i)k^2_i(t)+m_0\Pi(k_0)k_0^2(t)\label{eq8} \\
&\approx m^2(2t+m_0)^{\frac{1}{2}} \Big(\sum_{t_i=1}^t \frac{1}{(2t_i+m_0)^{\frac{3}{2}}}+m_0^{-\frac{1}{2}}\Big).
\end{align}
For large time, $\sum_{t_i=1}^t\Big(\frac{1}{t_i}\Big)^{\frac{3}{2}}=\zeta(\frac{3}{2}) \approx 2.612$, we obtain 
$<k^2_s(t)> \approx m^2\sqrt{2t}(m_0^{-\frac{1}{2}}+2.612-b)$
with $b=1+\frac{1}{2^\frac{3}{2}}+\frac{1}{3^\frac{3}{2}}+\ldots+\frac{1}{(1+m_0)^\frac{3}{2}}$.\\
Fluctuations of $<k_s(t)>$ are given by
 \begin{align}
(\Delta k_s(t))^2 \equiv <k_s(t)^2>-<k_s(t)>^2\approx m^2 \Big[(m_0^{-\frac{1}{2}}+2.612-b)\sqrt{2t}-(\ln(2t))^2\Big],
\label{eq10}
\end{align}
which become arbitrary large when time increases sufficiently.\\
Simulation data, in accordance with Eq.~\eqref{eq10} (see Fig.~\ref{fig2}(b)), shows the increasing tendency of 
fluctuations in $<~k_s(t)>$.
This can be explained by the fact that the maximum degree in the network $k_{max}\sim \sqrt{t}$ increases \cite{CO} 
faster than  $<k_s(t)>\sim \ln(t)$ (Eq.~\eqref{eq7}) 
and the difference between the two quantities becomes greater with time.\\
We now turn to the same analysis in our model.
The mean field evolution equation for $k_i(t)$ gives
\begin{eqnarray}
 \dfrac{\partial k_i(t)}{\partial t}+\dfrac{k_i(t)}{(2t+m_0)(t+m_0-1)}=\dfrac{m}{t+m_0-1}.
\end{eqnarray}
The solution has the form
\begin{eqnarray}
k_i(t)=m \Big(\frac{t+m_0-1}{2t+m_0}\Big)^{\frac{1}{m_0-2}}\Bigg[\Big(\frac{t_i+m_0-1}{2t_i+m_0}\Big)^{-\frac{1}{m_0-2}}-A(t_i)
+A(t)\Bigg],
\label{kit}
\end{eqnarray}
where $A(t)=\displaystyle \int_1^{t}\dfrac{\Big(\frac{t'+m_0-1}{2t'+m_0}\Big)^{-\frac{1}{m_0-2}}}{t'+m_0-1}dt'$.\\
The average value of target node $<k_s(t)>$ is obtained immediately for any time $t$ by substituting Eq.~\eqref{kit}
 into Eq.~\eqref{eq6}. The resulting equation is solved numerically as shown in Fig.~\ref{fig3}(a).\\
For large time and taking $t\gg m_0$, we find
$A(t)\approx 2^{\frac{1}{m_0-2}} \ln(t)$, $k_i(t)\approx m\Big(1+\ln \dfrac{t}{t_i}\Big)$, then
\begin{align}
 \label{kom}
<k_s(t)>&\approx\dfrac{m}{t}\Big(\sum_{t_i=1}^{t} \ln(t)-\ln(t_i)+1\Big)\\
&\approx\dfrac{m}{t}\Big(t \Big(\ln(t)+1\Big)-\Big(\sum_{t_i=1}^{t} \ln(t_i)\Big)\Big) \nonumber \\
&\approx\dfrac{m}{t}\Big(t\ \Big(\ln(t)+1\Big)-\ln(t_i!)\Big) \nonumber \\
&\approx2m \nonumber
\end{align}
This is the value of the node average connectivity
of a network growing with uniform attachment resulting in an exponential distribution of degree \cite{BJ}. \\
The second moment is obtained by substituting the corresponding expressions of $\Pi(k_i)$ and $k_i(t)$ in Eq.~\eqref{eq8}, 
we get for large time
\begin{eqnarray}
\begin{split}
<k^2_s(t)>\approx \frac{m^2}{t}\Big(\sum_{t_i=1}^t (\ln(\frac{t}{t_i})+1)^2\Big).
\label{k2om}
\end{split}
\end{eqnarray}
Making the approximations $\sum_{t_i=1}^t \ln(t_i)\approx t\ln(t)-t$, and
 $\sum_{t_i=1}^t \ln(t_i)^2\approx t\ln(t)^2-2t\ln(t)+2t-2$, we find
$<k^2_s(t)>\approx 5m^2.$\\
Fluctuations are $(\Delta k_s(t))^2 \equiv {<k^2_s(t)>-<k_s(t)>^2}\approx m^2$.
This finding, together with $<k_s(t)>\approx 2m$, show that almost all nodes have the same degree as illustrated in
 Fig.~\ref{fig3}(b). The homogeneity of the network can be explained by the fact that the preferential attachment used here 
doesn't allow the formation of hubs, since it neither allows the rich to get richer, nor it enriches the poor.  

\section{Conclusion}
In this work, we have introduced a simple model of complex network with a preferential
attachment criteria  and without "rich get richer" effect.  The network obtained is
homogeneous, which demonstrates the crucial role of the "rich get richer" in the topology
of the network.
Giving preferential treatment to the least connected nodes is equivalent to use a random attachment probability.
In terms of social wealth distribution, Pareto principle \cite{PA} doesn't apply and we have instead an 
exponential distribution of income.\\
Computing the instantaneous average degree of a target node and its fluctuations provide more information
than the usual average degree of the network, in particular we show how the average degree of hubs and its
fluctuations diverge with time in the BA model, and stay finite in our model.

\end{document}